\documentclass{aa}  
\usepackage{graphicx}
\usepackage{amsmath}
\usepackage{amssymb}
\usepackage{txfonts}
\usepackage{xcolor}

\newcommand{\ros}{{\it ROSAT}}

\newcommand{\xmm}{{\it XMM-Newton}}
\newcommand{\eROS}{eROSITA}
\newcommand{\swift}{{\it Swift}}

\newcommand{\nh}{N_{\rm H}}
\def \msev{M7}
\def \magoe{\object{RX~J1856.5-3754}}
\def \magzs{\object{RX~J0720.4-3125}}
\def \magos{\object{RX~J1605.3+3249}}
\def \magot{\object{RX~J1308.6+2127}}

\def \magze{\object{RX~J0806.4-4123}}
\def \magzf{\object{RX~J0420.0-5022}}
\def \carINS{\object{2XMM~J104608.7-594306}}

\def \insc{J0221}
\def \jztto{\object{4XMM~J022141.5-735632}} 
\begin{document} 
\title{XMM-Newton and SRG/\eROS\ observations of the isolated neutron star candidate \jztto
\thanks{Based on observations obtained with \xmm, an ESA science mission with instruments and contributions directly funded by ESA Member States and NASA (observations 0884190401, 0674110401).}} 
\author{A.~M.~Pires\inst{1}
   \and C.~Motch\inst{2}
   \and J.~Kurpas\inst{1}
   \and A.~D.~Schwope\inst{1}
   \and F.~Valdes\inst{3}
   \and F.~Haberl\inst{4}
   \and I.~Traulsen\inst{1}
   \and D.~Tub\'in\inst{1}
   \and W.~Becker\inst{4}
   \and J.~Comparat\inst{4}
   \and C.~Maitra\inst{4}
   \and A.~Meisner\inst{3}
   \and J.~Moustakas\inst{5}
   \and M.~Salvato\inst{4}
}
\offprints{A. M. Pires}
\institute{Leibniz-Institut f\"ur Astrophysik Potsdam (AIP), An der Sternwarte 16, 14482 Potsdam, Germany
   \email{apires@aip.de} 
   \and
   CNRS, Universit\'e de Strasbourg, Observatoire Astronomique, 11 rue de l'Universit\'e, F-67000 Strasbourg, France
   \and
   NSF's National Optical/Infrared Research Laboratory (NOIRLab), 950 N. Cherry Ave, Tucson, AZ 85732, USA
   \and
   Max-Planck-Institut f\"ur extraterrestrische Physik, Giessenbachstra\ss e, 85748 Garching, Germany
   \and
   Department of Physics \& Astronomy, Siena College, 515 Loudon Road, Loudonville, NY, 12211, USA
}
\date{Received ...; accepted ...}
\titlerunning{The isolated neutron star candidate \jztto}
\authorrunning{A.~M.~Pires et al.}
\abstract
{We report the results of follow-up investigations of a possible new thermally emitting isolated neutron star (INS), \jztto, using observations from \xmm\ and {\it Spectrum Roentgen Gamma} (SRG) \eROS. The analysis is complemented by Legacy Survey imaging in the optical and near-infrared wavelengths. The X-ray source, the first to be targeted by \xmm\ in an effort to identify new INS candidates from the fourth generation of the \xmm\ serendipitous source catalogue Data Release~9 (4XMM-DR9), shows a remarkably soft energy distribution and a lack of catalogued counterparts; the very high X-ray-to-optical flux ratio virtually excludes any other identification than an INS. Within current observational limits, no significant flux variation nor change of spectral state is registered over nearly ten years. Future dedicated observations, particularly to search for pulsations, are crucial to shed further light on the nature of the X-ray source and relations to other Galactic neutron stars.}
\maketitle
\section{Introduction\label{sec_intro}}
The vast majority of the neutron stars in our Galaxy are observed at radio wavelengths. Yet, it is arguably in X-rays that isolated neutron stars (INSs) reveal their diversity in all their complexity \citep[see, e.g.][for an overview]{2010PNAS..107.7147K}. In particular, the group of radio-quiet thermally emitting INSs dubbed the ``magnificent seven'' (\msev, for simplicity), originally identified in the \ros\ All-Sky Survey data as soft, bright X-ray sources (observed fluxes typically within $\sim10^{-12}-10^{-11}$\,erg\,s$^{-1}$\,cm$^{-2}$) with no obvious optical counterparts, share a rather well-defined set of properties that have never been encountered together in previously known families of INSs \citep[see][for reviews]{2007Ap&SS.308..181H,2009ASSL..357..141T}. 
The \msev\ are locally as numerous as young radio and $\gamma$-ray pulsars and may belong to a formerly neglected component of the overall Galactic INS population. The discovery of similar sources beyond our local volume is therefore key to understanding their properties as a group and relations to other families of Galactic INSs \citep[e.g.][]{2010MNRAS.401.2675P,2013MNRAS.434..123V}.

At fainter X-ray fluxes ($f_{\rm X}\lesssim1.5\times10^{-13}$\,erg\,s$^{-1}$\,cm$^{-2}$), source confusion and contamination from other classes of X-ray emitters hamper the identification of new members due to the large positional and spectral uncertainties. In preparation for the full sensitivity of the \eROS\ All-Sky Survey \citep[eRASS;][]{2021A&A...647A...1P}, the use of serendipitous data from the \xmm\ Observatory \citep{2020A&A...641A.136W} provides an excellent opportunity to test search algorithms and discover new INSs beyond the solar vicinity.

Building on our experience cross-correlating previous releases of the \xmm\ catalogue of X-ray sources \citep[][]{2009A&A...504..185P, 2017ASPC..512..165M}, we searched the source content of 4XMM-DR9 for new INS candidates. We used as main criteria the absence of any catalogued optical/UV/IR counterpart and a soft spectrum down to a limiting flux of $10^{-14}$\,erg\,s$^{-1}$\,cm$^{-2}$ in the $0.5-1$\,keV energy band (see Section~\ref{sec_selectiondr9}, for details). Among the selected sources, we retrieve known thermally emitting INSs: the \msev, quiescent neutron stars in globular clusters, and our ``own'' cooling INS identified in the Carina Nebula, \carINS\ \citep[see][and references therein]{2015A&A...583A.117P}. The brightest among the unknown sources were then selected for dedicated observations with \xmm\ (fulfil programs 088419, 090126; \citeauthor{pires+dr9}, in preparation).

We report the first results of this follow-up effort on \jztto, an X-ray source located in the direction of the Magellanic Bridge and independently identified by similar searches in 4XMM-DR10 and eRASS data (\citealt{2022MNRAS.509.1217R}; \citeauthor{kurpas+erass}, private communication). 
The overall properties of \jztto, as obtained from the 4XMM-DR9 catalogue, are 
listed in Table~\ref{tab_4xmm}.
\begin{table*}[t]
\caption{Properties of the target from 4XMM-DR9.
\label{tab_4xmm}}
\centering
\begin{tabular}{lrrccrrrrr}
\hline\hline
Target & RA & Dec & Error & $N_{\rm H}^{\rm Gal}$\,\tablefootmark{$\star$} & $b$ & Flux\,\tablefootmark{$\dagger$} & HR$_1$\,\tablefootmark{$\ddag$} & HR$_2$\,\tablefootmark{$\ddag$} & HR$_3$\,\tablefootmark{$\ddag$}\\
4XMM & (\degr) & (\degr) & (\arcsec) & (cm$^{-2}$) & (\degr) & & & & \\
\hline
J022141.5-735632 & $35.42309$ & $-73.94223$ & $0.9$ & $1.41\times10^{21}$ & $-41.7$ & $9.8(6)$ & $-0.663(21)$ & $-0.91(4)$ & $-0.95(24)$\\ 
\hline
\end{tabular}
\tablefoot{
\tablefoottext{$\star$}{Total hydrogen column density in the line-of-sight \citep{2016A&A...594A.116H}.}
\tablefoottext{$\dagger$}{The catalogued EPIC flux is in units of $10^{-14}$\,erg\,s$^{-1}$\,cm$^{-2}$ in the $0.2-12$\,keV energy band.}
\tablefoottext{$\ddag$}{Hardness ratios (HR) are ratios of the difference to total counts in the five \xmm\ energy bands, $0.2-0.5$\,keV, $0.5-1$\,keV, $1-2$\,keV, $2-4.5$\,keV, and $4.5-12$\,keV. The source signal is undefined in HR$_4$.}}
\end{table*}
In the following, we present the \xmm\ and \eROS\ data set that forms the backbone of the analysis; we show that the X-ray emission of the target is predominantly thermal and constant over many years; the absence of optical counterparts in Legacy Survey\footnote{\texttt{https://legacysurvey.org}} \citep{2019AJ....157..168D} Data Release~10 imaging safely excludes a more ordinary X-ray emitter than an INS. We finalise with a discussion of our state of knowledge and interpretation on the nature of the source, with prospects for future investigations.
\section{Observations and data reduction\label{sec_obs}}
\subsection{XMM-Newton\label{sec_obsxmm}}
\begin{table}
\small
\caption{X-ray observations used in the analysis.
\label{tab_obs}}
\centering
\begin{tabular}{lccrr}
\hline\hline
\textit{Obsid}, Sky field & Date & MJD & Exposure & GTI \\
 & & (days) & (s) & (\%)\\
\hline
\multicolumn{5}{l}{\xmm}\\
\hline
0674110401\tablefootmark{$\star$} & 2012-02-09 & $55\,967.076$ & $27\,543$ & $93$\\
0884190401\tablefootmark{$\dagger$} & 2021-07-09 & $59\,405.013$  & $28\,010$ & $73$\\
\hline
\multicolumn{5}{l}{\eROS\tablefootmark{$\ddag$}}\\
\hline
03716501 & 2020-04-25 & $58\,970.200$ & $1\,008$ & $81$\\
03716502 & 2020-10-28 & $59\,155.966$ & $936$ & $82$\\
03716503 & 2021-04-28 & $59\,338.242$ & $1\,037$ & $78$\\
03716504 & 2021-10-30 & $59\,523.091$ & $950$ & $74$\\
\hline
\end{tabular}
\tablefoot{The EPIC cameras operated in full-frame imaging mode with
\tablefoottext{$\star$}{medium and}
\tablefoottext{$\dagger$}{thin filters.}
\tablefoottext{$\ddag$}{All seven \eROS\ telescope modules (``TM0'') were operated in SURVEY mode with FILTER setup.}}
\end{table}
The \xmm\ Observatory \citep{2001A&A...365L...1J} targeted the X-ray source \jztto\ (\insc, for short) for circa 50\,ks on July 9--10 2021. 
We included in the analysis the only other \xmm\ observation that serendipitously detected the target in 2012 (0674110401). 
We performed standard data reduction with SAS~20 (xmmsas\_20211130\_0941), using the most up-to-date calibration files and following the analysis guidelines of each EPIC instrument. We filtered the event lists to exclude ``bad'' CCD pixels and columns and retain the photon patterns with the highest quality energy calibration. The source centroid and optimal extraction region in each EPIC camera, with typical sizes of $30\arcsec-40$\arcsec, were defined with the SAS task \texttt{eregionanalyse} in the $0.2-2$\,keV energy band; the X-ray emission of the source is compatible with the background level at energies above 2\,keV. Background circular regions of size 100\arcsec\ were defined away from the source, on the same CCD of the target whenever possible. 
We show in Table~\ref{tab_obs} the instrumental setup, net exposure, and percentage of good-time intervals (GTIs) of the \xmm\ observations included in the analysis. 

The parameters of the target (the detection likelihood, counts, rates, and hardness ratios, as extracted from a maximum likelihood PSF fitting on the EPIC images of each \xmm\ epoch) are listed in the two first columns of Table~\ref{tab_psfparams}. The parameters are determined with the SAS task \texttt{edetect\_chain} on images created for each camera, observation, and standard catalogue energy bands (see caption of Table~\ref{tab_4xmm}, for a definition). Only the combined EPIC results are shown.
\begin{figure}[t]
\begin{center}
\includegraphics[width=\linewidth]{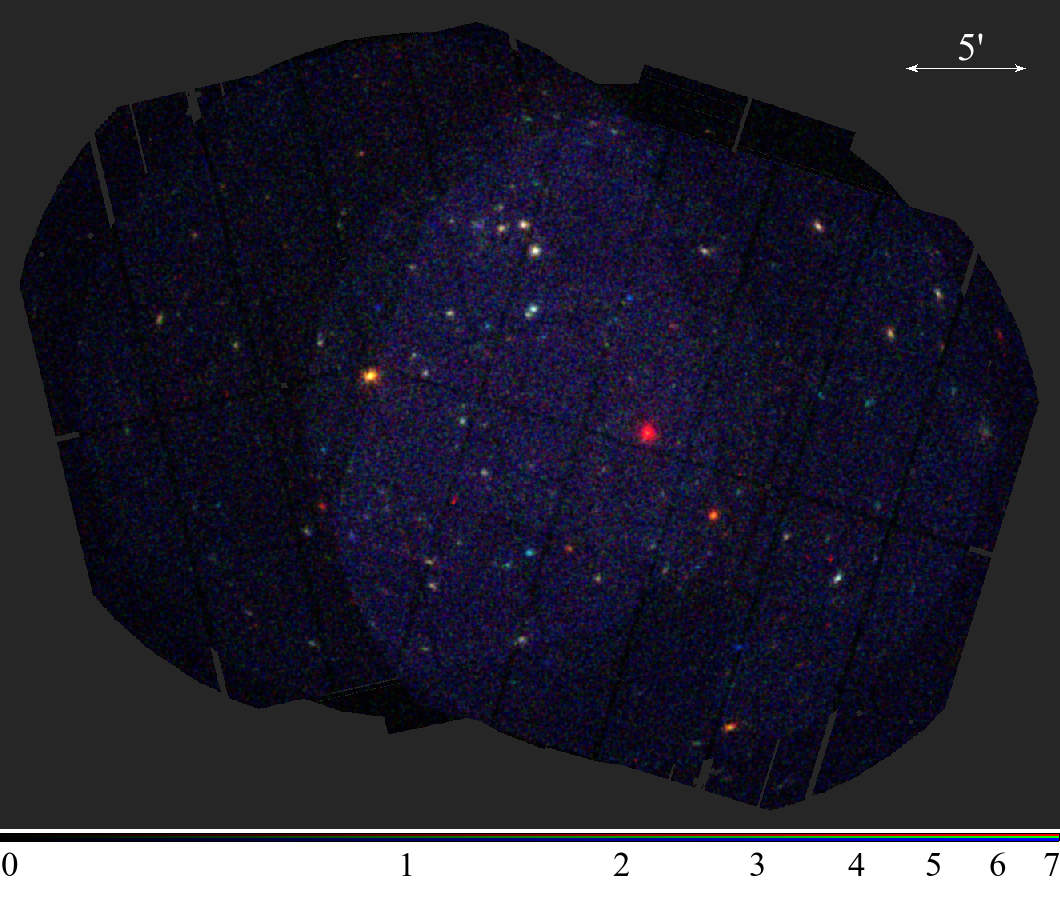}\vskip1pt
\includegraphics[width=\linewidth]{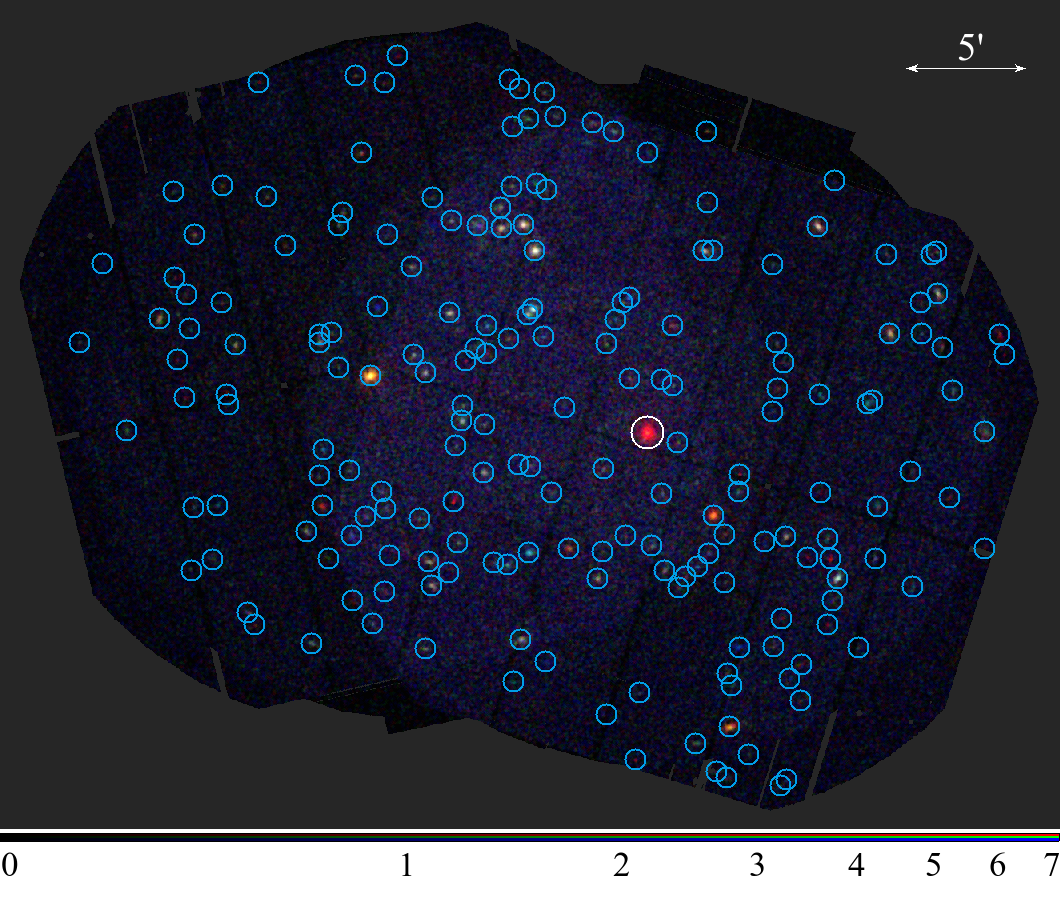}
\caption{Smoothed false-colour (\textit{top}) and source detection (\textit{bottom}) images of the two \xmm\ observations covering the position of \insc. The red band covers the energies from 0.2$-$1.0\,keV, green from 1.0$-$2.0\,keV, and blue from 2.0$-$12.0\,keV. The position of the target and of other sources detected in the field-of-view are marked by white and light blue circles, respectively.\label{fig_stack}}
\end{center}
\end{figure}
\begin{table}[t]
\small
\caption{Target parameters from PSF fitting (\xmm).
\label{tab_psfparams}}
\centering
\begin{tabular}{@{}lrrr@{}}
\hline\hline
& 0674110401 & 0884190401 & stack \\
\hline
Off-axis (\arcmin) & $10$ & $0$ & $0$ \\ 
$\mathcal{L}$\,\tablefootmark{$\star$} & $3500$ & $5350$ & $11\,675$\\ 
Counts & $1\,310(40)$ & $2\,110(50)$ & $4\,184(70)$ \\ 
Rate (s$^{-1}$)  & $0.0909(29)$ & $0.0896(23)$ & $0.1081(18)$ \\
HR$_1$ & $-0.628(24)$ & $-0.576(21)$ & $-0.657(13)$ \\ 
HR$_2$ & $-0.90(5)$ & $-0.92(3)$ & $-0.93(2)$ \\ 
HR$_3$ & $-0.90(50)$ & $-0.90(50)$ & $-0.81(26)$ \\ 
RA (J2000)\,\tablefootmark{$\dagger$}  & $02\ \ 21\ \ 41.52$ & $02\ \ 21\ \ 41.55$ & $02\ \ 21\ \ 41.57$\\ 
Dec (J2000)\,\tablefootmark{$\dagger$} & $-73\ \ 56\ \ 33.70$ & $-73\ \ 56\ \ 33.80$ & $-73\ \ 56\ \ 34.10$\\ 
Error (\arcsec) & $0.27$ & $0.20$ & $0.10$ \\
RA offset\,\tablefootmark{$\ddag$} (\arcsec)  & $-0.46(29)$ & $-1.03(27)$ & $-0.20(30)$\\
Dec offset\,\tablefootmark{$\ddag$} (\arcsec)  & $-0.51(28)$ & $-1.26(27)$ & $-0.10(30)$\\
Ref.~sources & $76$& $139$ & $150$\\
\hline
\end{tabular}
\tablefoot{The parameters of the source in each column are determined with the meta-tasks \texttt{edetect\_chain} and \texttt{edetect\_stack} (EPIC, $0.2-12$\,keV). Errors are $1\sigma$ statistical uncertainties. 
\tablefoottext{$\star$}{Likelihood of detection.}
\tablefoottext{$\dagger$}{Coordinates of the target, corrected for the astrometry with the SAS task \texttt{eposcorr}.}
\tablefoottext{$\ddag$}{Positional offsets in each coordinate determined for the astrometry (see text; for details).}}
\end{table}
We used the SAS task \texttt{eposcorr} to refine the astrometry by cross-correlating the list of EPIC X-ray source positions with those of optical Guide Star Catalogue \citep{2021yCat.1353....0L} objects lying within 15\arcmin\ around the nominal pointing coordinates. The small positional offsets in right ascension and declination listed in Table~\ref{tab_psfparams} were consistently detected with the Gaia EDR3 and AllWISE catalogues \citep{2020yCat.1350....0G, 2014yCat.2328....0C}. The agreement between epochs is good overall ($<3\sigma$).
\begin{figure}[t]
\begin{center}
\includegraphics[width=0.495\textwidth]{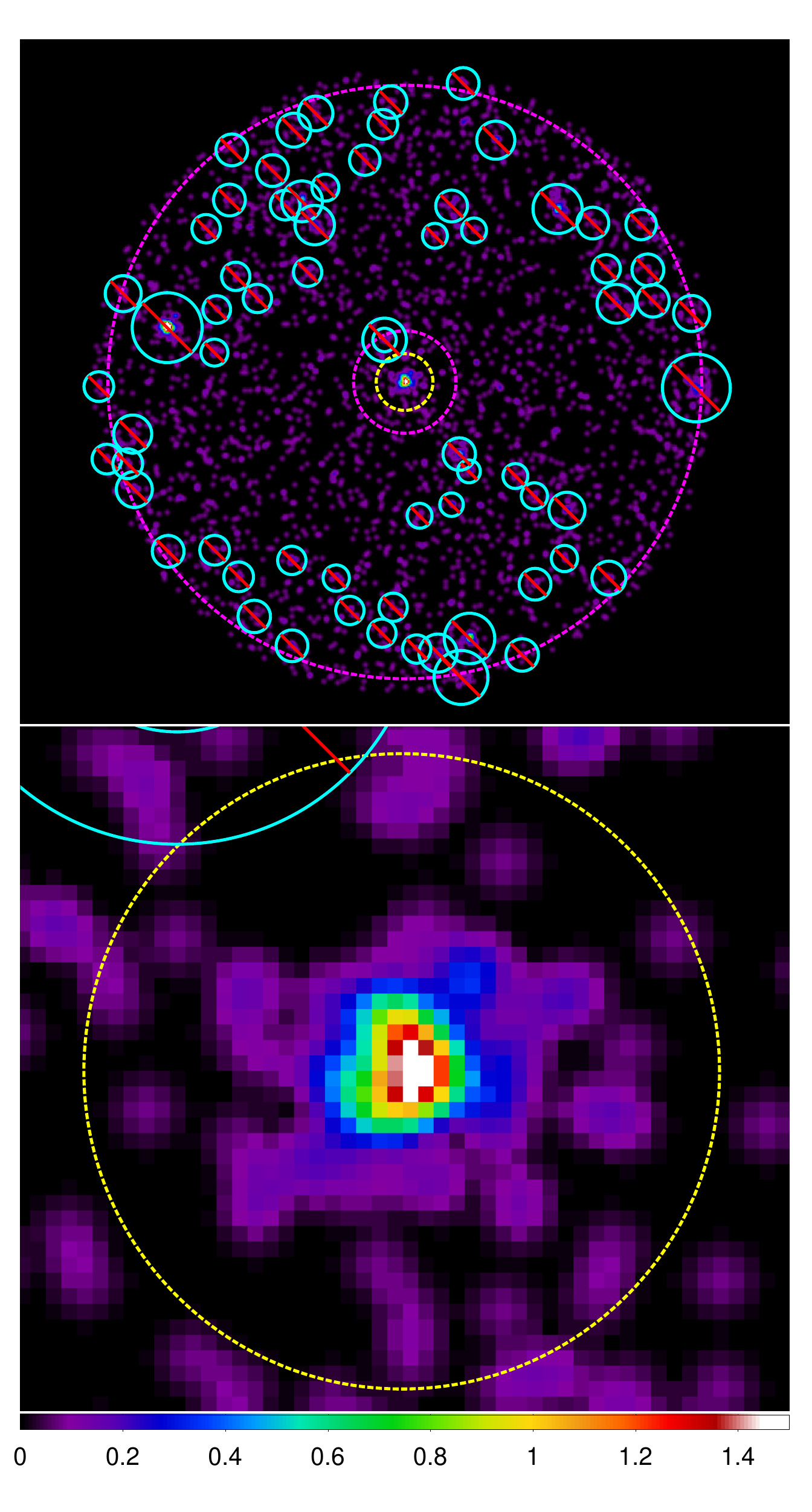}
\caption{\eROS\ extraction regions (eRASS:4, TM0, $0.2-1$\,keV). \textit{Top.} Background annulus (magenta) with inner and outer radii of sizes 2.5\arcmin\ and 15\arcmin, respectively, excluding ``contaminants'' (cyan circles with a red bar across). \textit{Bottom.} Source circular region (yellow) of size 82\arcsec.\label{fig_erassreg}}
\end{center}
\end{figure}

Stacked source detection on both \xmm\ observations was performed using the \xmm\ SAS task \texttt{edetect\_stack} \citep{2019A&A...624A..77T} and the 4XMM-DR12s catalogue pipeline. First, the results of the \texttt{eposcorr} astrometric correction of the individual observations were applied to the attitude information in order to shift the events to rectified positions, as described by \citet{2020A&A...641A.137T}.
We then extracted images, background maps, exposure maps, and detection masks for both observations with reference coordinates centred at the position of \insc. Deviating from the default parameters, we used an image binning of 2\arcsec$\times$2\arcsec\ pixels and a PSF fit radius of 2\arcmin\ about the sources, in order to optimise the sensitivity to faint sources in these fields, which are dominated by point-like sources and do not suffer from source confusion, and in order to achieve optimum positional accuracy of the target. Maximum-likelihood PSF source detection was run simultaneously in the five \xmm\ energy bands.
In the whole field, 199 sources are detected with a minimum detection likelihood of six in the stack or in at least one contributing observation. Figure~\ref{fig_stack} shows a false-colour mosaic image of the field, colour coding energies between 0.2 and 1\,keV (red), 1\,keV and 2\,keV (green), and 2\,keV and 12\,keV (blue). For displaying purposes, it has been smoothed by a top-hat function with a two-pixel radius. The X-ray soft \insc\ is prominently visible in the red band.
\subsection{\eROS\label{sec_obseros}}
The All-Sky Survey of \eROS, the first to be performed at X-ray energies since the \ros\ era \citep{1999A&A...349..389V}, is expected to surpass its predecessor's sensitivity by a factor of $25$ \citep[$0.2-2.3$\,keV;][]{2021A&A...647A...1P}.
Since the beginning of the survey in December 2019, \eROS\ has successfully completed four of the planned eight all-sky scans. The eRASS catalogues are created by and made available to the members of the German \eROS\ Consortium. We searched the individual (eRASS1--4) and stacked (eRASS:4) X-ray source catalogues for unidentified INSs above a limiting flux of $10^{-13}$\,erg\,s$^{-1}$\,cm$^{-2}$ in the $0.2-2$\,keV energy band (\citeauthor{kurpas+erass}, in preparation).
Having survived our probabilistic catalogue cross-matching and screening procedure (only about $0.05\%$ of over $10^5$ X-ray sources satisfy all the selection criteria for identification with an INS candidate), \insc\ is one of a handful of intrinsically soft point sources with no obvious counterparts that have been lined up for follow-up investigations in the optical and in X-rays.

The \eROS\ instrument consists of seven telescope modules (TM 1--7) operating in the $0.2-10$\,keV energy band. The detectors of TM 5 and 7 (``TM 9'') are more sensitive to soft X-rays, as they lack the aluminium on-chip optical light filter the other five cameras (``TM8'') carry. 
The TM9 detectors are known to suffer from time-variable light leaks that affect their performance and calibration at the softest energies \citep[see][for details]{2021A&A...647A...1P}.

For the analysis, we retrieved the event lists corresponding to the target's sky tile in each eRASS1--4. We list in Table~\ref{tab_obs} an overview of these observations. The position of the X-ray source was repeatedly covered in April and October 2020--2021, with exposures of $\sim0.9-1$\,ks.
The data sets were processed with pipeline version c020 and analysed in the $0.2-10$\,keV energy band with the eSASS software system, applying up-to-date calibration files (eSASSusers\_211214; \citealt{2022A&A...661A...1B}). 
For all observations, the event files of each individual TM were filtered for periods of high background activity with the eSASS task \texttt{flaregti} in the $2.2-10$\,keV energy band. Unless otherwise noted, all valid photon patterns and active telescope modules (``TM0'') were considered for optimal sensitivity. 
The total sum of GTIs accumulated over the four epochs is 2.3\,ks ($\sim1$\,ks, corrected for the vignetting). Altogether, $101\pm12$ photons are collected from the X-ray source in the $0.2-1$\,keV energy band, considering all seven active TMs.

Similarly to the \xmm\ observations, we analysed the X-ray source content of the observations and the parameters of the target using maximum likelihood PSF fitting \citep{2022A&A...661A...1B}. 
Based on the results of source detection and PSF fitting we used the ``auto'' option of the eSASS task \texttt{srctool} to create optimised source and background extraction regions over the cumulative eRASS:4 events (Fig.~\ref{fig_erassreg}). The positions of the 67 X-ray sources detected in the field-of-view in the $0.2-1$\,keV energy band were excluded from the background region.
Finally, we verified the statistics of the eRASS light curves of the target for general trend variability. 
The light curves were corrected for bad pixels, vignetting, exposure, and background counts with the eSASS task \texttt{srctool}. The $0.2-1$\,keV count rate of the target is consistent with a constant value.
\section{Results\label{sec_analysis}}
\subsection{Timing analysis}
We searched for periodic signals that could be associated with the rotation period of the X-ray source (hundreds of milliseconds to tens of seconds); alternatively, to identify flux modulations of the order of tens of seconds to hours as observed in novae and supersoft X-ray sources \citep{2022ApJ...932...45O}. We considered only the two $\sim30$\,ks \xmm\ observations of \insc\ (Table~\ref{tab_obs}), the individual eRASS exposures being too shallow for a meaningful timing analysis. For maximum sensitivity, we included all valid patterns and events of the three EPIC cameras unfiltered for GTIs in the $0.2-2$\,keV energy band.
The times-of-arrival of the photons, barycentred to the rest frame of the solar system using the SAS task \texttt{barycen} and the source coordinates (Table~\ref{tab_psfparams}), were Fourier-analysed in frequency domain to search for the presence of periodic signals \citep{1983A&A...128..245B}. We analysed the EPIC cameras together in the $\Delta\nu=(0.002-1.9)\times 10^{-1}$\,Hz frequency range; the higher time resolution of the pn camera with respect to MOS allows one to search the pn time series in a broader frequency range, up to $\sim6.8$\,Hz, albeit with somewhat lower sensitivity. We adopted in both (EPIC/pn) searches frequency steps of $8-10$\,$\mu$Hz, which correspond to an oversampling factor of $\sim3$. The number of statistically independent trials are $\sim(2-3)\times10^5$ and $\sim(6-8)\times10^3$, respectively, for pn and EPIC searches in the 2012 and 2021 observations.
We found no statistically significant ($>4\sigma$) periodicity in neither epoch. The most constraining $3\sigma$ upper limits on the source pulsed fraction, $p_{\rm f}^{\rm pn}<13$\% ($P\sim0.15-5000$\,s) and $p_{\rm f}^{\rm EPIC}<10$\% ($P\sim5.2-5000$\,s), are derived from the 2021 observation.
\subsection{Spectral analysis and long-term variability\label{sec_spec}}
\begin{table}
\small
\caption{Results of spectral fitting.
\label{tab_spec}}
\centering
\begin{tabular}{@{}lcrrrrr@{}}
\hline\hline
 & NHP\tablefootmark{a} & $C$\,(d.o.f.) & $\nh$\,\tablefootmark{b} & \multicolumn{2}{c}{Model parameters} & $F_X$\,\tablefootmark{c}\\
\hline
\multicolumn{4}{l}{\texttt{bbodyrad}} & $kT$ & $R_{\rm em}$\,\tablefootmark{d} & \\
 & & & & (eV) & (km) & \\
\hline
1 & 57 & 162\,(159) & $7.2_{-0.8}^{+0.9}$ & $60.2(1.7)$ & $8.7_{-1.2}^{+1.4}$ & $5.7_{-0.8}^{+1.0}$\\
2\tablefootmark{$\star$} & 72 & $141\,(157)$ & $8.8_{-1.0}^{+1.2}$ & $55.7_{-2.7}^{+2.3}$ & $12.7_{-2.4}^{+3.7}$ & $8.1_{-1.4}^{+2.0}$\\
3\tablefootmark{$\dagger$} & 73 & $141\,(157)$ & $9.3_{-1.2}^{+1.6}$ & $54_{-4}^{+3}$ & $14_{-3}^{+5}$ & $9.2_{-1.9}^{+2.9}$\\
4 & 49 & 157\,(151) & $7.2_{-0.8}^{+0.9}$ & $62(4)$ & $11(4)$ & $5.7_{-0.8}^{+1.0}$\\
\hline
\multicolumn{4}{l}{\texttt{nsa}\tablefootmark{e}} & $T_{\rm eff}$ & $d$ & \\
 & & & & ($10^5$\,K) & (pc) & \\
\hline
5 & 27 & 175\,(156) & $8.6_{-0.3}^{+0.7}$ & $1.91_{-0.04}^{+0.05}$ & $28_{-3}^{+4}$ & $15.9_{-2.6}^{+3.8}$\\
6 & 63 & 155\,(156) & $10.5_{-0.8}^{+1.1}$ & $2.65_{-0.16}^{+0.13}$ & $43_{-11}^{+10}$ & $14.5_{-2.2}^{+3.5}$\\
7 & 63 & 155\,(156) & $10.4(9)$ & $3.00(15)$ & $68_{-15}^{+17}$ & $9.5_{-0.7}^{+1.3}$\\
\hline
\end{tabular}
\tablefoot{Errors are $1\sigma$ confidence levels. Models fitted to the data: (1) single- and (2) double-temperature blackbody; (3) blackbody plus power-law tail, (4) ``variable'' single-temperature blackbody; (5) non-magnetised neutron star atmosphere, and neutron star atmosphere with magnetic field intensities (6) $10^{12}$\,G and (7) $10^{13}$\,G. 
\tablefoottext{a}{Null-hypothesis probability in percentage that the data are drawn from the model.}
\tablefoottext{b}{The column density is in units of $10^{20}$\,cm$^{-2}$.}
\tablefoottext{c}{The unabsorbed flux is in units of $10^{-13}$\,erg\,s$^{-1}$\,cm$^{-2}$ ($0.2-12$\,keV).}
\tablefoottext{d}{The blackbody emission radius at infinity is computed assuming a distance of $1$\,kpc.}
\tablefoottext{e}{The \texttt{nsa} models assume canonical neutron star mass and radius, 1.4\,M$_{\odot}$ and 10\,km.}
\tablefoottext{$\star$}{Parameters of the best-fit second blackbody component: $kT_{\rm hot}=190_{-50}^{+70}$\,eV and $R_{\rm em}^{\rm hot}=0.7_{-0.4}^{+1.0}$\,km.}
\tablefoottext{$\dagger$}{Parameters of the power-law tail: photon index $\Gamma=4.2_{-1.3}^{+1.6}$ and observed flux $f_{\rm X}^{\rm PL}=(6.1_{-1.4}^{+1.5})\times10^{-14}$\,erg\,s$^{-1}$\,cm$^{-2}$ ($0.2-2$\,keV).}}
\end{table}
The analysed data set comprises six epochs, ten spectra\footnote{Specifically, 2 EPIC + 4 eRASS epochs and 2 $\times$ (pn, MOS1, MOS2) + 4 $\times$ TM0 spectra.
}, and over $4\,000$ counts ($0.2-2$\,keV); background noise amounts to up to $15$\% and $30$\%, respectively, of the total EPIC and \eROS\ events. 
The spectral analysis was restricted to GTI-filtered photons, single- and double-pattern events for pn, and all valid CCD patterns for the MOS and TM cameras.
We regrouped the energy channels to avoid low ($<5$) counts per spectral bin and kept oversampling of the instrumental resolution capped to a maximum factor of 3. 
To fit the spectra we used XSPEC 12.10.1f and the Cash C statistic \citep{1996ASPC..101...17A,1979ApJ...228..939C}; unless otherwise noted, the fit parameters were allowed to vary freely within reasonable ranges. The spectra were fitted simultaneously allowing for a renormalisation factor to account for cross-calibration uncertainties between epochs and instruments. We verified that the inclusion of photons from \eROS\ detectors 5 and 7 (those affected by optical leaks; see Section~\ref{sec_obseros}) do not bias the parameter estimation of the results reported here. Finally, we adopted the photoionization cross-sections of \citet{1996ApJ...465..487V}, the photoelectric absorption model \texttt{tbabs}, and elemental abundances of \citet{2000ApJ...542..914W} to account for the interstellar material in the line-of-sight; while testing for variable elemental abundances, we adopted the absorption model \texttt{tbvarabs} and checked the improvement in the fit statistic element-wise by means of a F-test. Complimentary information can be found in Sections~\ref{sec_variability} and \ref{sec_additspec}.

\begin{figure}[t]
\begin{center}
\includegraphics[width=0.49\textwidth]{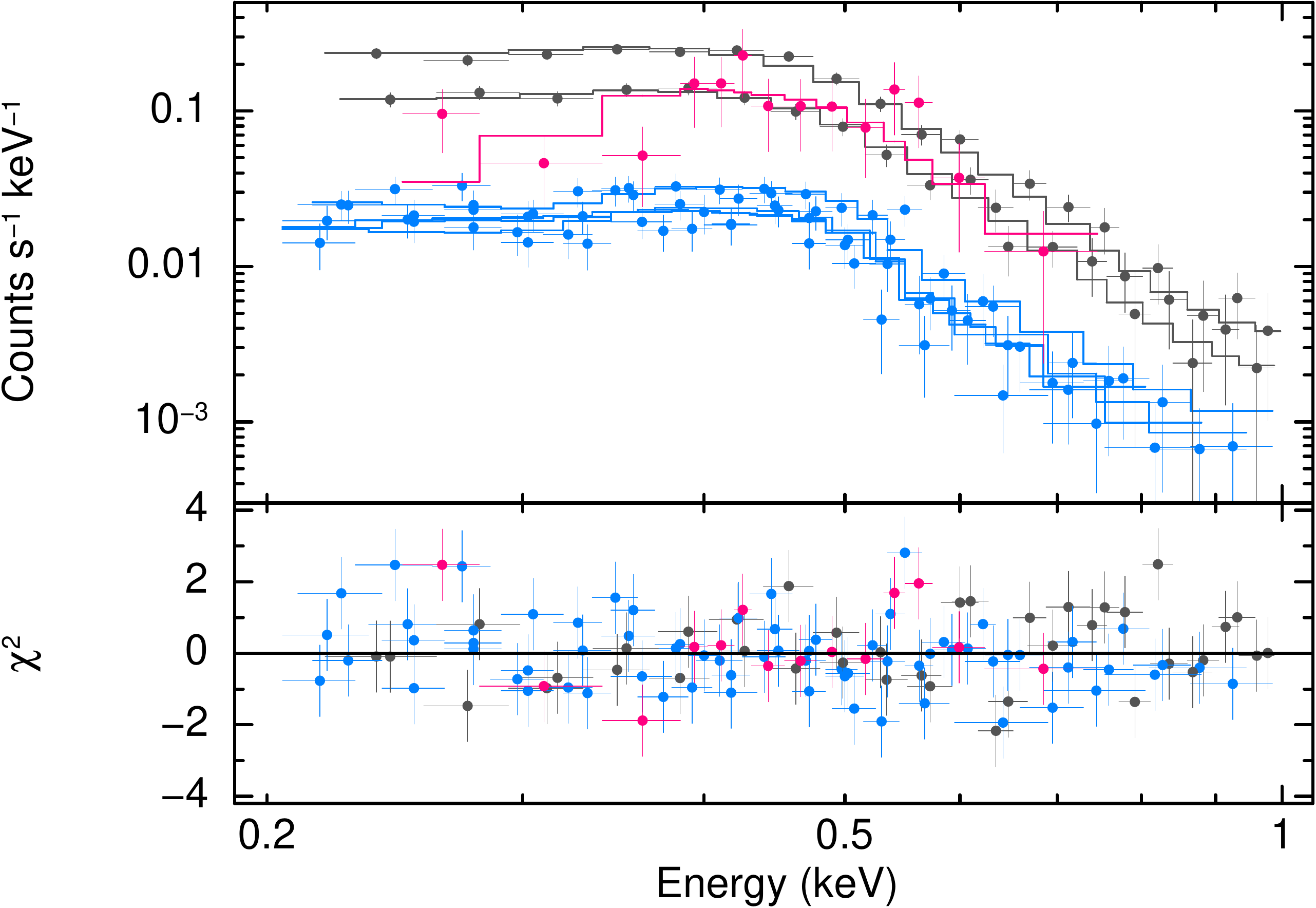}
\caption{Best-fit double-blackbody model and residuals (fit 2 in Table~\ref{tab_spec}). The two EPIC pn and four MOS data sets are colour-coded in dark grey and blue, respectively. The \eROS\ spectrum (merged here for display purposes) is shown in pink.\label{fig_specfit}}
\end{center}
\end{figure}

The main results of the spectral analysis are summarised in Table~\ref{tab_spec}.
In model (1) we assumed a single-temperature absorbed blackbody and constancy between all observations. The best solution consists of a soft blackbody with $kT=60.2\pm1.7$\,eV, absorbed by a column density of $\nh=7.2_{-0.8}^{+0.9}\times10^{20}$\,cm$^{-2}$; the observed flux is $f_{\rm X}=1.18(3)\times10^{-13}$\,erg\,s$^{-1}$\,cm$^{-2}$ ($0.2-2$\,keV) and the size of the emission region, at a distance of 1\,kpc, is $8.7_{-1.2}^{+1.4}$\,km. 
The inclusion of an additional component -- either a second blackbody with $kT=140-260$\,eV or a power-law tail with photon index $\Gamma=4.2_{-1.3}^{+1.6}$ -- is statistically significant (F-test probability of $1.8\times10^{-5}$; see models 2 and 3 in Table~\ref{tab_spec}). In both cases, there are no significant changes ($>2\sigma$) to the dominant blackbody component, nor to $\nh$. 
In Fig.~\ref{fig_specfit} we show the spectral data folded with model (2) and residuals.

\begin{figure*}[t]
\begin{center}
\includegraphics[width=0.49\textwidth]{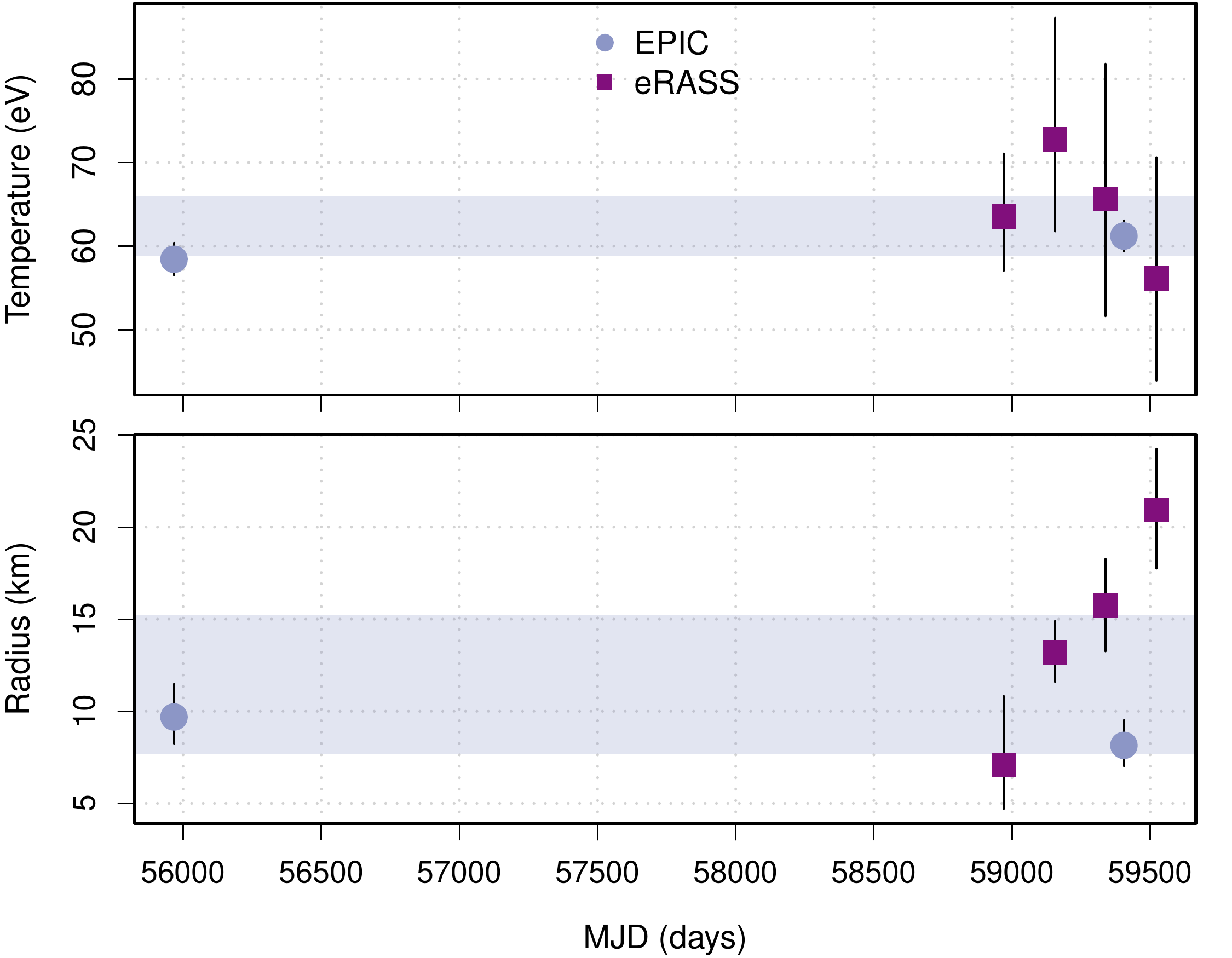}\hfill
\includegraphics[width=0.49\textwidth]{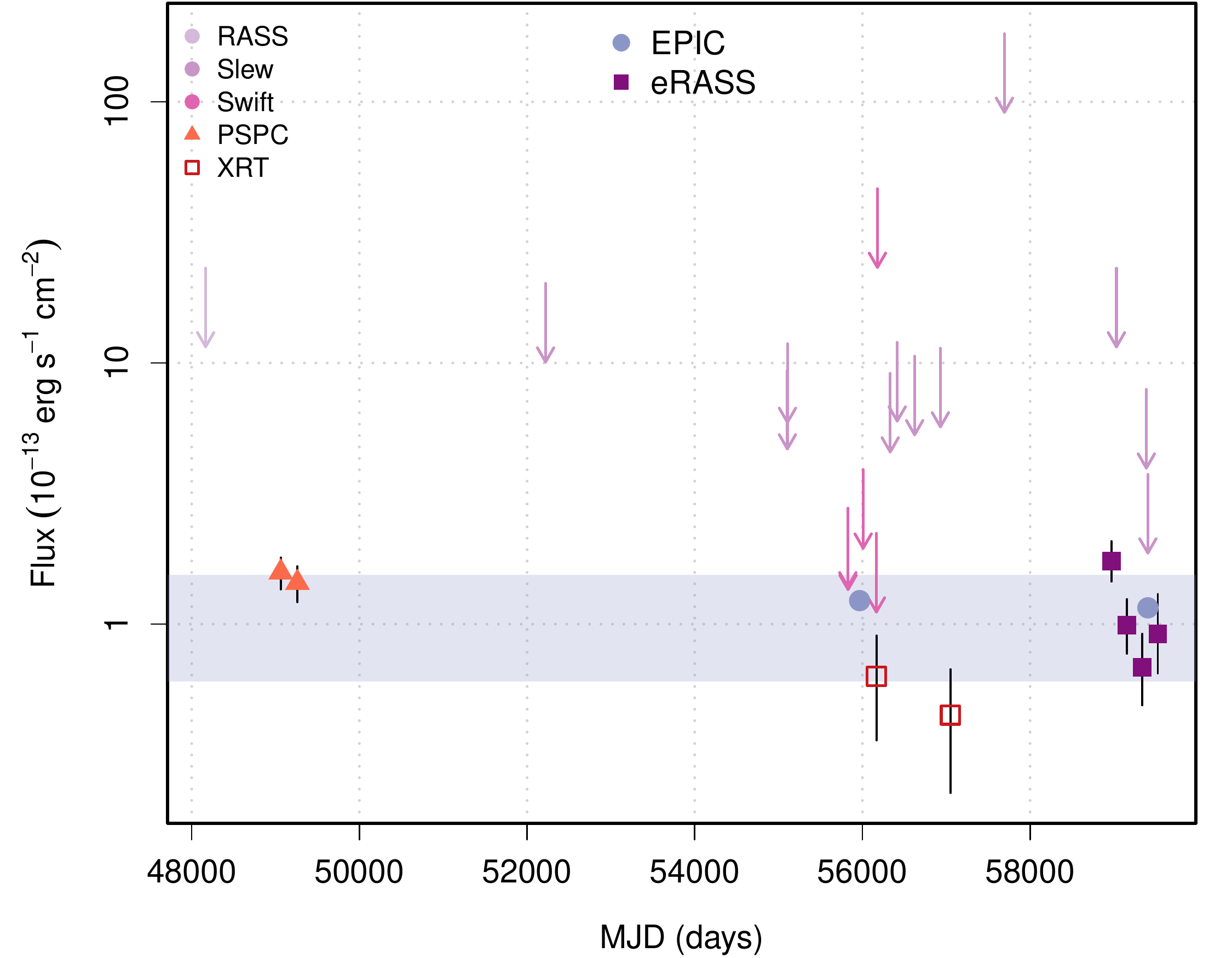}
\caption{\textit{Left.} Blackbody temperature and emission radius as a function of MJD (``variable'' model 4; see Tables~\ref{tab_spec} and \ref{tab_specepoch}). \textit{Right.} Long-term evolution of the $0.2-2$\,keV X-ray flux of the target, including upper limits and previous detection by other X-ray missions. The time interval extends back to the \ros\ All-Sky Survey (RASS) and pointed (PSPC) era and include data points from Swift XRT and \xmm\ slew observations (see the text, for details). In all plots the purple horizontal shaded areas show the $1\sigma$ median absolute deviation of the parameters.
\label{fig_fluxspecevol}}
\end{center}
\end{figure*}

In model (4) we allowed the parameters of model (1) except for $\nh$ to assume independent values in each epoch. We list in Table~\ref{tab_spec} the median and median absolute deviation of the parameters; see Table~\ref{tab_specepoch} and Fig.~\ref{fig_fluxspecevol}, for the individual values. We find no evidence of long-term variability: the deviations from a constant value are well within the expected cross-calibration uncertainties between the EPIC instruments \citep[$2$\% in $kT$ and up to $3$\% in flux;][]{2014A&A...564A..75R}; the source parameters agree within the 90\% confidence level and the fit has a non-acceptable F-test probability with respect to model (1). The much wider spread and trends observed in \eROS\ data (left panel of Fig.~\ref{fig_fluxspecevol}), discussed in Section~\ref{sec_variability}, may be asserted to low count statistics and remaining calibration issues at the softest end of the observatory passband. 
\begin{table}[t]
\small
\caption{Investigation of inter-epoch variability.
\label{tab_specepoch}}
\centering
\begin{tabular}{@{}ccrrrrr@{}}
\hline\hline
MJD & Data set & Counts & $\mathcal{B}$\tablefootmark{a} & $kT$ & $R_{\rm em}$\tablefootmark{b} & $f_{\rm X}$\tablefootmark{c}\\
(days) & & & (\%) & (eV) & (km) & \\
\hline
$55967$ & EPIC  & $1596$ & $12$ & $58.4_{-1.9}^{+2.0}$ & $9.7_{-1.4}^{+1.8}$  & $1.23_{-0.04}^{+0.05}$\\
$58970$ & eRASS & $43$   & $12$ & $64_{-7}^{+8}$       & $7.1_{-2.4}^{+4.0}$  & $1.74_{-0.29}^{+0.30}$\\
$59156$ & eRASS & $25$   & $17$ & $73_{-11}^{+15}$     & $13.2_{-2.6}^{+2.8}$ & $0.99_{-0.22}^{+0.25}$\\
$59338$ & eRASS & $20$   & $28$ & $66_{-14}^{+16}$     & $16(4)$              & $0.68_{-0.19}^{+0.23}$\\
$59405$ & EPIC  & $2679$ & $13$ & $61.2_{-1.8}^{+1.9}$ & $8.1_{-1.1}^{+1.4}$  & $1.15_{-0.03}^{+0.03}$\\
$59523$ & eRASS & $20$   & $30$ & $56_{-12}^{+15}$     & $21_{-5}^{+6}$       & $0.92_{-0.27}^{+0.40}$\\
\hline
\end{tabular}
\tablefoot{Errors are $1\sigma$ confidence levels. The model fitted to the data is (4) of Table~\ref{tab_spec} (``variable'' single-temperature blackbody). 
\tablefoottext{a}{Percentage of background in each data set ($0.2-2$\,keV).}
\tablefoottext{b}{The blackbody emission radius at infinity is computed assuming a distance of $1$\,kpc.}
\tablefoottext{c}{The observed flux is in units of $10^{-13}$\,erg\,s$^{-1}$\,cm$^{-2}$ ($0.2-2$\,keV).}}
\end{table}

Interestingly, the analysis of the long-term evolution of the $0.2-2$\,keV X-ray flux of \insc, including upper limits and previous detection by other X-ray missions, suggests that the properties of the X-ray source are stable over 30 years (Fig.~\ref{fig_fluxspecevol}; right).
In addition to the EPIC and eRASS observations analysed here, we employed the web-based tool High-Energy Lightcurve Generator \cite[HILIGT\footnote{\texttt{http://xmmuls.esac.esa.int/upperlimitserver}};][]{2022A&C....3800531S} to retrieve upper limits and flux values for the source. These are determined through aperture photometry at \insc's sky position and the Bayesian approach of \citet{1991ApJ...374..344K} on archival observations from \swift\ XRT, \ros, and the \xmm\ Slew Survey \citep[see][and references therein]{2022MNRAS.511.4265R}. We assumed an absorbed blackbody spectral model consistent with the source's spectral shape and selected a $3\sigma$ confidence interval to estimate upper limits (single-sided 99.7$\%$ probability). The measurements are overall consistent with a constant flux.
\begin{figure}[t]
\begin{center}
\includegraphics[width=0.475\textwidth]{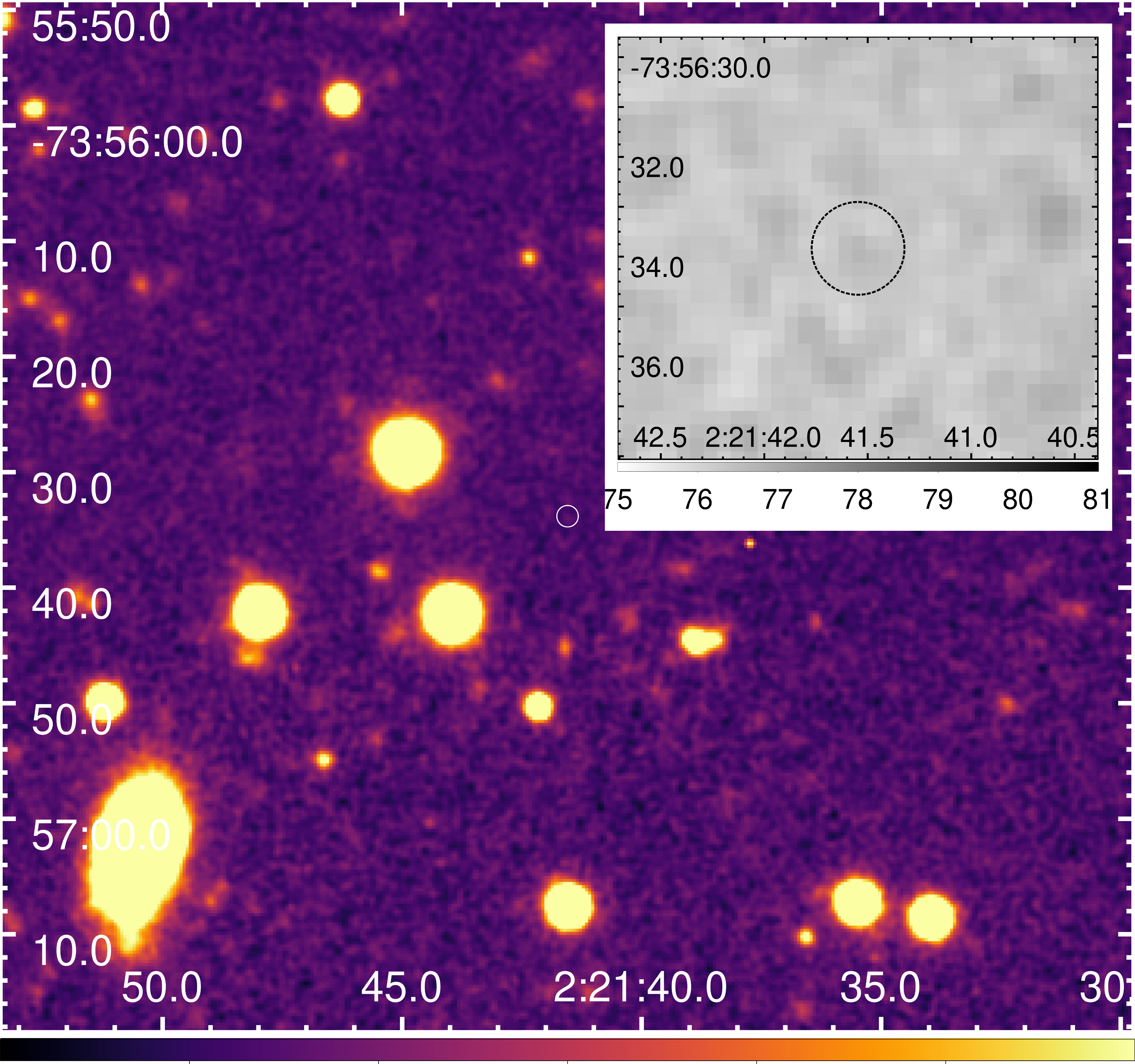}
\caption{Photometric-weighted multi-filter stack of Legacy Survey DR10 $g, r, z$ and DECam $u, i$ images centred on the sky position of \insc. The blank ($m_{\rm G}>26.45$) $3\sigma$ confidence level error circle of the X-ray source is displayed in the inset in an inverted black and white colour map.} \label{deep_stack}
\end{center}
\end{figure}

Neutron star atmosphere models, with $B=10^{12}-10^{13}$\,G, $\nh=1.1\times10^{21}$\,cm$^{-2}$ and $T_{\rm eff}\sim(2.7-3)\times10^5$\,K, describe the spectrum of the source as well as the multi-component models, and can be considered an improved fit with respect to the non-magnetic case (models 5--7 in Table~\ref{tab_spec}). Assuming emission from the entire stellar surface, canonical neutron star mass and radius, and a fully ionized hydrogen atmosphere (\texttt{nsa} in XSPEC; \citealt{1995ASIC..450...71P}) the model normalisation requires in all cases a nearby neutron star within 90\,pc ($3\sigma$). These may be rather unrealistic assumptions for the atmosphere of non-accreting neutron stars, given the large increase of ionization potential expected under such conditions of temperature and magnetic field \citep[see][for a comprehensive review]{2016arXiv160501281P}. Other spectral models did not provide compelling results. In particular, the temperature of the best-fit partially ionized neutron star atmosphere (\texttt{nsmaxg} in XSPEC; \citealt{2008ApJS..178..102H}) is outside the computed model grid, $\log(T_{\rm eff})<5.5$, while the photon index of the best-fit power-law model is unreasonably steep ($\Gamma\sim8$). We note that our results are in perfect agreement with the findings of \citet{2022MNRAS.509.1217R}.
\subsection{Optical and near-infrared limits}
We see no evidence in the optical and near-infrared for the source in the Legacy Survey Data Release~10 (DR10). The survey follows the same data reduction of Data Release~9 \citep{2019AJ....157..168D}, but covers almost entirely the German part of the \eROS\ sky\footnote{The \eROS\ data rights are equally split between a German and a Russian Consortium; see \citet[][for details]{2021A&A...647A...1P}.} at the depth of the Dark Energy Survey in {\it g, r, i, z} \citep{2018ApJS..239...18A}. The depth and the addition of the {\it i} band increases the number of detected sources by 30\% (Legacy Survey Collaboration 2022, private communication). In the infrared the surveys makes use of 8 years of NEOWISE observations \citep{2014ApJ...792...30M}. To search for the maximum limit of the available DECam archival imaging at this sky location, additional exposures in $u$ and $i$ were added and a depth weighted, multi-filter stack of the 32 exposures with exposure times greater than 60\,s was produced by the DECam Community Pipeline \citep{2014ASPC..485..379V}. Figure~\ref{deep_stack} is the resultant deep stack with the position of the INS candidate indicated. Clearly, there is no sign of a source in this blank region. The depth limit is obtained from the local measurement of the background noise (standard deviation of 0.52 data numbers), point-spread full-width at half-maximum (4.23\,pixels corresponding to 1.06\arcsec), and zero-point calibration to Gaia-EDR G-band (28.68). Using the formula for the depth in an optimal Gaussian aperture with the measured values yields a G-band depth limit of 26.45 (25.89) at $3\sigma$ ($5\sigma$). These allow us to derive a $3\sigma$ lower limit on the target's X-ray-to-optical flux ratio, $\log(F_\mathrm{X}/F_\mathrm{V})\gtrsim3.7$. 
We assumed a flat spectrum to translate the G-band flux to that in the V-band and adopted the best-fit $F_{\rm X}$ and $\nh$ from model (2) in Table~\ref{tab_spec}; optical de-reddening and the total extinction specific to the Gaia-EDR G-band were computed with the usual relations of \citet{1995A&A...293..889P} and \citet{1989ApJ...345..245C}.
\section{Discussion and outlook\label{sec_disc}}
We report the first results of a campaign to follow-up INS candidates from 4XMM-DR9.
The X-ray source \jztto\ was put forward by \citet{2022MNRAS.509.1217R} on the same premise of a soft spectrum and lack of obvious counterparts; it is also a ``target of interest'' on dedicated searches in \eROS\ All-Sky Survey data.
The joint analysis of the \xmm\ and \eROS\ observations performed between 2012 and 2021 confirms the source's essentially thermal energy distribution, with $kT\sim60$\,eV, $\nh\sim7\times10^{20}$\,cm$^{-2}$, and $f_{\rm X}\sim1.2\times10^{-13}$\,erg\,s$^{-1}$\,cm$^{-2}$ ($0.2-2$\,keV). 
The optical limit derived from the deep stacked Legacy Survey DR10 and additional DECam images, $m_{\rm G}>26.45$ ($3\sigma$) in the Gaia-EDR G-band, excludes a cataclysmic variable or hot white dwarf in the foreground of the Magellanic Bridge. 

We find no evidence for variability in either flux or spectral state within the nearly ten-year interval covered by the analysis. Previous detection at a similar flux level by \ros\ PSPC and the \swift\ X-ray Telescope suggest that the emission is fairly stable over decade-long time scales. Additional monitoring with higher quality data will be necessary to exclude the few-percent level of spectral variation as reported for the M7 INSs \magzs\ and \magos\ \citep[]{2004A&A...415L..31D,2022arXiv220206793M}.

In addition to the main thermal component, excess emission above 0.7\,keV may be accommodated with either a second (hot) blackbody with $kT_{\rm hot}=190_{-50}^{+70}$\,eV or a power-law tail; for the latter, the photon index $\Gamma=4.2_{-1.3}^{+1.6}$ is considerably steeper than what is typically observed in the spectra of middle-aged spin-powered pulsars dominated by thermal emission \citep[e.g.][]{2022A&A...661A..41S,2022MNRAS.tmp.1097R}. 
Magnetised and fully ionized neutron star atmosphere models, with $B=10^{12}-10^{13}$\,G and $T_{\rm eff}=(2.6-3)\times10^5$\,K, are also in reasonable agreement with the data. All best-fit \texttt{nsa} models in Table~\ref{tab_spec} predict a rather close-by neutron star within 90\,pc ($3\sigma$). We caution, however, against the applicability of such simple model atmospheres for sources as soft as \insc\ when not in the weak-field regime \citep{2016arXiv160501281P}.

\begin{figure}[t]
\begin{center}
\includegraphics[width=0.475\textwidth]{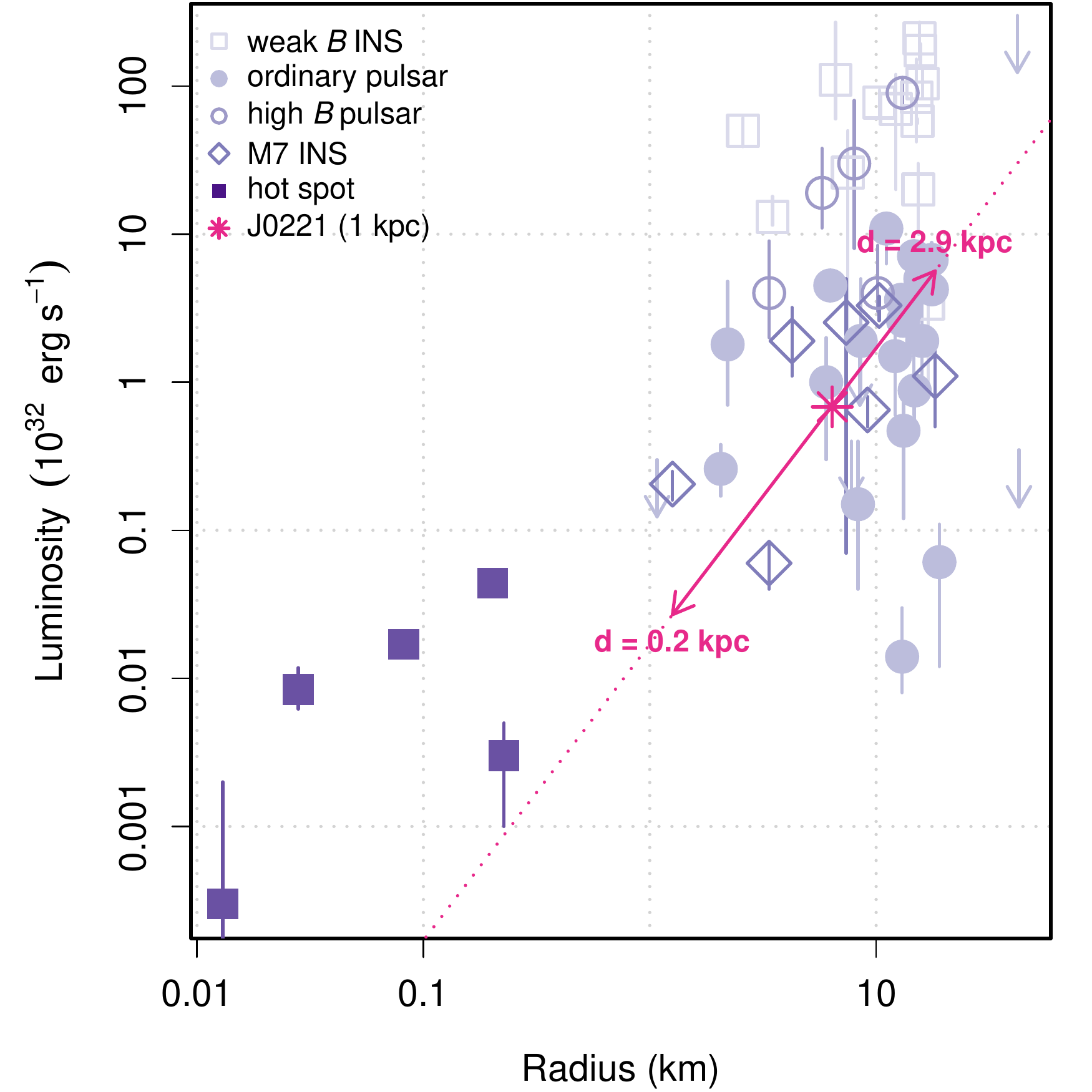}
\caption{Thermal luminosity \textit{vs.} emission radius for cooling neutron stars. The arrow shows the distance-luminosity range of the target assuming the same emission radii observed for the \msev\ INSs (see the text, for details).} \label{fig_rl}
\end{center}
\end{figure}

In Fig.~\ref{fig_rl} we show the range of luminosity and emission radius of \insc\ in the context of the sample of 55 cooling INSs from \citet[][we refer to their review paper for the terminology, references, and a description of the included data sets and spectral models]{2020MNRAS.496.5052P}. The distance interval of 200\,pc to 2.9\,kpc, indicated by the arrow in the diagram, assumes that the X-ray source has a similar emission radius as the \msev. Considering a fiducial timescale of 1.2\,Myr to cool down the surface to its present-day temperature of $\sim(3-7)\times10^5$\,K \citep[see, e.g.][their Fig.~1]{2004ARA&A..42..169Y}, the location of \insc\ $\sim100$\,pc to 1.9\,kpc below the Galactic disk ($b=-41.7$\degr; Table~\ref{tab_4xmm}) implies a projected speed within $90$\,km\,s$^{-1}$ and $1600$\,km\,s$^{-1}$, roughly in agreement with the range observed for radio pulsars \citep{2005MNRAS.360..974H}. 

From all our tested spectral models we infer a hydrogen column density a factor of $\sim1.5-3$ higher than the Galactic value from \citet[][$\nh=3.8\times10^{20}$\,cm$^{-2}$]{1990ARA&A..28..215D}, but within that derived from the HI4PI map in this line-of-sight \citep[$\nh=1.4\times10^{21}$\,cm$^{-2}$;][]{2016A&A...594A.116H}. The latter integrates the HI emission over both the Galactic and the Magellanic Cloud systems \citep[see also][]{2022A&A...661A..37S}. 
If the absorption is purely interstellar, this would place \insc\ outside the Milky Way into the Magellanic Bridge, at a distance of the order of that of the Small Magellanic Cloud, 60\,kpc. The inferred size of the emission area for such a large distance is incompatible with the neutron star surface. On the other hand, part of the absorption could be caused by matter local to the X-ray source \citep[e.g.~the case of \magze;][]{2018ApJ...865....1P}. Interestingly, residuals around $540-580$\,eV may indicate an enhancement of oxygen local to or in the direction of the INS candidate (Section~\ref{sec_additspec}). 
An updated value of the HI emission in this line-of-sight excluding the contribution of the Magellanic system is necessary for a reliable estimate of the amount of absorption intrinsic to the X-ray source.

X-ray emitting INSs at faint fluxes will consist of the products of more distant massive star associations and will be younger and hotter than the \msev\ on average \citep[see][for details based on a population synthesis model and a discussion of multiwavelength follow-up strategies of new INS candidates]{2017AN....338..213P}. The most compelling candidates will be located in the Galactic plane or close to open stellar clusters and supernova remnants. Other interesting Galactic and extragalactic X-ray sources (high-redshift quasars, ultraluminous and supersoft X-ray sources, millisecond and rotation-powered pulsars) may also be selected by our searches (\citeauthor[][in preparation]{pires+dr9}; see also \citealt{2022Univ....8..354K,2022A&A...658A..95V}). In particular, the sky location and softness of \insc\ could indicate a supersoft nature \citep[][for a review]{1997ARA&A..35...69K}. 
However, at the distance of the SMC, the luminosity of $\sim 2\times 10^{35}$\,erg\,s$^{-1}$ is well below the range for stable hydrogen burning typical of white dwarves in close binary systems \citep{2013ApJ...777..136W}. The size of the inferred emission region, $\sim290-410$\,km, is likewise inconsistent with the white dwarf radius. Given the absence of significant variability and optical counterparts, we regard the surpersoft source interpretation as unlikely.

Overall, the identification of \jztto\ with an INS appears robust. 
A systematic search for pulsations from dedicated searches in X-rays and in the radio regime is crucial to establish the true nature of the X-ray source and shed light on its relations to other Galactic neutron stars.
\begin{acknowledgements}
We thank the anonymous referee for useful comments and suggestions which helped to improve the paper.
This work was supported by the project XMM2ATHENA, which has received funding from the European Union's Horizon 2020 research and innovation programme under grant agreement n$^{\rm o}101004168$. IT gratefully acknowledges the support of SSC work at AIP by Deutsches Zentrum f\"ur Luft- und Raumfahrt (DLR) through grant 50\,OX\,1901. DT acknowledges support by DLR grants FKZ 50\,OR\,2203.
This research has made use of data obtained from the 4XMM XMM-Newton serendipitous source catalogue compiled by the 10 institutes of the XMM-Newton Survey Science Centre selected by ESA.
This work is based on data from \eROS\, the soft X-ray instrument aboard SRG, a joint Russian-German science mission supported by the Russian Space Agency (Roskosmos), in the interests of the Russian Academy of Sciences represented by its Space Research Institute (IKI), and the Deutsches Zentrum f\"ur Luft- und Raumfahrt (DLR). The SRG spacecraft was built by Lavochkin Association (NPOL) and its subcontractors, and is operated by NPOL with support from the Max Planck Institute for Extraterrestrial Physics (MPE).
The development and construction of the \eROS\ X-ray instrument was led by MPE, with contributions from the Dr. Karl Remeis Observatory Bamberg \& ECAP (FAU Erlangen-Nuernberg), the University of Hamburg Observatory, the Leibniz Institute for Astrophysics Potsdam (AIP), and the Institute for Astronomy and Astrophysics of the University of T\"ubingen, with the support of DLR and the Max Planck Society. The Argelander Institute for Astronomy of the University of Bonn and the Ludwig Maximilians Universit\"at Munich also participated in the science preparation for \eROS.
The \eROS\ data shown here were processed using the eSASS/NRTA software system developed by the German \eROS\ consortium.
The Legacy Surveys consist of three individual and complementary projects: the Dark Energy Camera Legacy Survey (DECaLS; Proposal ID \#2014B-0404; PIs: David Schlegel and Arjun Dey), the Beijing-Arizona Sky Survey (BASS; NOAO Prop. ID \#2015A-0801; PIs: Zhou Xu and Xiaohui Fan), and the Mayall z-band Legacy Survey (MzLS; Prop. ID \#2016A-0453; PI: Arjun Dey). DECaLS, BASS and MzLS together include data obtained, respectively, at the Blanco telescope, Cerro Tololo Inter-American Observatory, NSF’s NOIRLab; the Bok telescope, Steward Observatory, University of Arizona; and the Mayall telescope, Kitt Peak National Observatory, NOIRLab. The Legacy Surveys project is honored to be permitted to conduct astronomical research on Iolkam Du’ag (Kitt Peak), a mountain with particular significance to the Tohono O’odham Nation.
NOIRLab is operated by the Association of Universities for Research in Astronomy (AURA) under a cooperative agreement with the National Science Foundation.
This project used data obtained with the Dark Energy Camera (DECam), which was constructed by the Dark Energy Survey (DES) collaboration. Funding for the DES Projects has been provided by the U.S. Department of Energy, the U.S. National Science Foundation, the Ministry of Science and Education of Spain, the Science and Technology Facilities Council of the United Kingdom, the Higher Education Funding Council for England, the National Center for Supercomputing Applications at the University of Illinois at Urbana-Champaign, the Kavli Institute of Cosmological Physics at the University of Chicago, Center for Cosmology and Astro-Particle Physics at the Ohio State University, the Mitchell Institute for Fundamental Physics and Astronomy at Texas A\&M University, Financiadora de Estudos e Projetos, Fundacao Carlos Chagas Filho de Amparo, Financiadora de Estudos e Projetos, Fundacao Carlos Chagas Filho de Amparo a Pesquisa do Estado do Rio de Janeiro, Conselho Nacional de Desenvolvimento Cientifico e Tecnologico and the Ministerio da Ciencia, Tecnologia e Inovacao, the Deutsche Forschungsgemeinschaft and the Collaborating Institutions in the Dark Energy Survey. The Collaborating Institutions are Argonne National Laboratory, the University of California at Santa Cruz, the University of Cambridge, Centro de Investigaciones Energeticas, Medioambientales y Tecnologicas-Madrid, the University of Chicago, University College London, the DES-Brazil Consortium, the University of Edinburgh, the Eidgenossische Technische Hochschule (ETH) Zurich, Fermi National Accelerator Laboratory, the University of Illinois at Urbana-Champaign, the Institut de Ciencies de l’Espai (IEEC/CSIC), the Institut de Fisica d’Altes Energies, Lawrence Berkeley National Laboratory, the Ludwig Maximilians Universitat Munchen and the associated Excellence Cluster Universe, the University of Michigan, NSF’s NOIRLab, the University of Nottingham, the Ohio State University, the University of Pennsylvania, the University of Portsmouth, SLAC National Accelerator Laboratory, Stanford University, the University of Sussex, and Texas A\&M University.
The Legacy Survey team makes use of data products from the Near-Earth Object Wide-field Infrared Survey Explorer (NEOWISE), which is a project of the Jet Propulsion Laboratory/California Institute of Technology. NEOWISE is funded by the National Aeronautics and Space Administration.
\end{acknowledgements}
\bibliographystyle{aa}
\bibliography{ref_dr9}
\begin{appendix}
\section{Selection of INS candidates\label{sec_selectiondr9}}
To search for INS candidates we start with a preliminary screening to remove all spurious sources arising from soft strips that appear in some parts of CCD4 in MOS1. We then select sources with positions in the HR$_2$-HR$_3$ hardness ratio diagram consistent with those of the \msev\ group of X-ray thermally emitting INSs (see, e.g.~Fig.~1 of \citealt{2009A&A...504..185P} and similar hardness ratio diagrams in \citealt{2022MNRAS.509.1217R}; we refer to the caption of Table~\ref{tab_4xmm}, for a definition of hardness ratios and standard \xmm\ energy bands). Practically, this implies selecting sources with a very soft spectral slope in the $0.5-2$\,keV range and consistent with no detected flux above 2\,keV, namely: $\rm{HR}_2-\sigma_{\rm{HR2}}<-0.3$ and HR$_3-\sigma_{\rm{HR3}}<-0.99$. Since HR$_1$ mostly depends on interstellar absorption, we do not put any constraint on its value.  
In order to exclude optically bright classes of soft X-ray emitters such as active coronae or cataclysmic variables, we remove \xmm\ sources having a positional match with a probability higher than 50\% with a selection of archival catalogues \citep{2017A&A...597A..89P}. For this purpose, we use the results of the statistical cross-matching process between 4XMM-DR9 and GALEX GR6+7, XMM-OM-SUSS4.1, Gaia DR2, APASS, SDSSDR12, Pan-STARRS DR1, SkyMapper, 2MASS and AllWISE \citep[see][for a description]{2020A&A...641A.136W}. Visual screening allows us to discard high proper motion active coronae which cannot match the X-ray source position due to epoch difference. Finally, we look for possible identifications in the SIMBAD\footnote{\texttt{http://simbad.u-strasbg.fr/simbad}} astronomical database. 
\section{Inter-epoch variability in eRASS\label{sec_variability}}
The best-fit parameters resulting from the analysis of the \eROS\ epochs, when fitted independently of EPIC, have shown a strong dependence on the choice of TM detectors, patterns, and binning. For example, adopting a simple blackbody model, the best-fit temperature of the target derived from TM8 is 30\% softer than when TM 5 and 7 are included in the analysis. More consistent results between the \eROS\ detectors are obtained when $\nh$ is fixed to the best-fit EPIC value. In particular, the fit of a constant blackbody model on the \eROS\ data sets has a high null-hypothesis probability, 80\% for 19 degrees of freedom, and parameters in overall agreement with EPIC: $kT=67_{-6}^{+7}$\,eV, $R_{\rm em}=7_{-2}^{+3}$\,km and $f_{\rm X}=(1.15_{-0.19}^{+0.15})\times10^{-13}$\,erg\,s$^{-1}$\,cm$^{-2}$.

For model (4) in Table~\ref{tab_spec}, we only kept the column density towards the target and the parameters of the blackbody within a given epoch (that is, between the EPIC cameras) coupled during fitting. 
When the requirement of constancy is lifted from the individual epochs, the parameters of the target as derived from eRASS in model (4) show a much wider spread around the median in comparison to the EPIC data, with variations in temperature and flux of up to 13\% and 60\%, respectively (Table~\ref{tab_specepoch} and Fig.~\ref{fig_fluxspecevol}). The chi-square values assuming constancy are $3.6$, $13.2$, and $8.2$ (5 d.o.f.), respectively for $kT$, $R_{\rm em}$ and $f_{\rm X}$.
\begin{table}[t]
\small
\caption{\eROS\ All-Sky Survey observations of thermal INSs.
\label{tab_erassm7}}
\centering
\begin{tabular}{@{}lccccr@{}}
\hline\hline
Sky field & Date\tablefootmark{$\star$} & MJD & Exp. & GTI\tablefootmark{$\dagger$} & Rate\tablefootmark{$\ddag$} \\
 & & (days) & (s) & (\%) & (s$^{-1}$) \\
\hline
\multicolumn{6}{c}{\magzf} \\
\hline
06714101 & 2020-01-24 & $58\,877.77$ & $941$ & $78$ & $0.108(19)$ \\
06714102 & 2020-07-24 & $59\,060.03$ & $950$ & $80$ & $0.146(21)$ \\
06714103 & 2021-01-18 & $59\,236.45$ & $634$ & $80$ & $0.121(23)$ \\
06714104 & 2021-07-25 & $59\,424.41$ & $648$ & $78$ & $0.126(23)$ \\
\hline
\multicolumn{6}{c}{\magzs} \\
\hline
11112001 & 2020-04-21 & $58\,962.14$ & $245$ & $71$ & $6.67(29)$\\
11112002 & 2020-10-24 & $59\,147.77$ & $216$ & $80$ & $7.17(31)$\\
11112003 & 2021-04-23 & $59\,329.52$ & $302$ & $79$ & $7.42(25)$\\
11112004 & 2021-10-25 & $59\,514.40$ & $317$ & $87$ & $7.76(33)$\\
\hline
\multicolumn{6}{c}{\magze} \\
\hline
12113201 & 2020-05-07 & $58\,978.14$ & $331$ & $71$ & $1.58(13)$ \\
12113202 & 2020-11-07 & $59\,162.53$ & $288$ & $74$ & $1.55(12)$ \\
12113203 & 2021-05-10 & $59\,345.93$ & $259$ & $76$ & $1.74(13)$ \\
12113204 & 2021-11-09 & $59\,529.57$ & $288$ & $77$ & $1.79(14)$ \\
\hline
\multicolumn{6}{c}{\carINS}\\
\hline
16315001 & 2020-01-01 & $58\,852.22$ & $533$ & $84$ & $0.086(25)$\\
16315002 & 2020-07-01 & $59\,034.39$ & $576$ & $84$ & $0.085(25)$\\
16315003 & 2021-01-01 & $59\,217.67$ & $418$ & $81$ & $0.111(33)$\\
16315004 & 2021-07-04 & $59\,404.56$ & $799$ & $91$ & $0.096(32)$\\
\hline
\multicolumn{6}{c}{\magot} \\
\hline
19606901 & 2019-12-15 & $58\,834.01$ & $245$ & $73$ & $3.02(19)$ \\
19606902 & 2020-06-14 & $59\,015.18$ & $187$ & $79$ & $2.67(21)$ \\
19606903 & 2020-12-17 & $59\,202.22$ & $230$ & $82$ & $2.62(19)$ \\
19606904 & 2021-06-19 & $59\,385.89$ & $245$ & $77$ & $2.46(17)$ \\
\hline
\multicolumn{6}{c}{\magoe}\\
\hline
28212901 & 2020-04-09 & $58\,949.34$ & $115$ & $69$ & $5.86(31)$\\
28212902 & 2020-10-13 & $59\,136.08$ & $115$ & $69$ & $5.48(30)$\\
28212903 & 2021-04-08 & $59\,313.30$ & $158$ & $73$ & $5.60(30)$\\
28212904 & 2021-10-10 & $59\,498.62$ & $173$ & $80$ & $5.53(27)$\\
\hline
\end{tabular}
\tablefoot{All \eROS\ observations were processed with pipeline version c020. The telescope modules (TMs) were operated in SURVEY mode with FILTER setup. 
\tablefoottext{$\star$}{Start date of a given eRASS epoch (of duration $\sim2-11$ days) at the sky location of the target.}
\tablefoottext{$\dagger$}{The percentage of good-time intervals are averaged over all active TMs.}
\tablefoottext{$\ddag$}{We show the count rate of the targets from source detection and PSF fitting over the TMs unaffected by optical light leaks, with $1\sigma$ errors ($0.2-1$\,keV; see Fig.~\ref{fig_ratesm7}).}
}
\end{table}

\begin{figure*}[t]
\begin{center}
\includegraphics[width=0.95\textwidth]{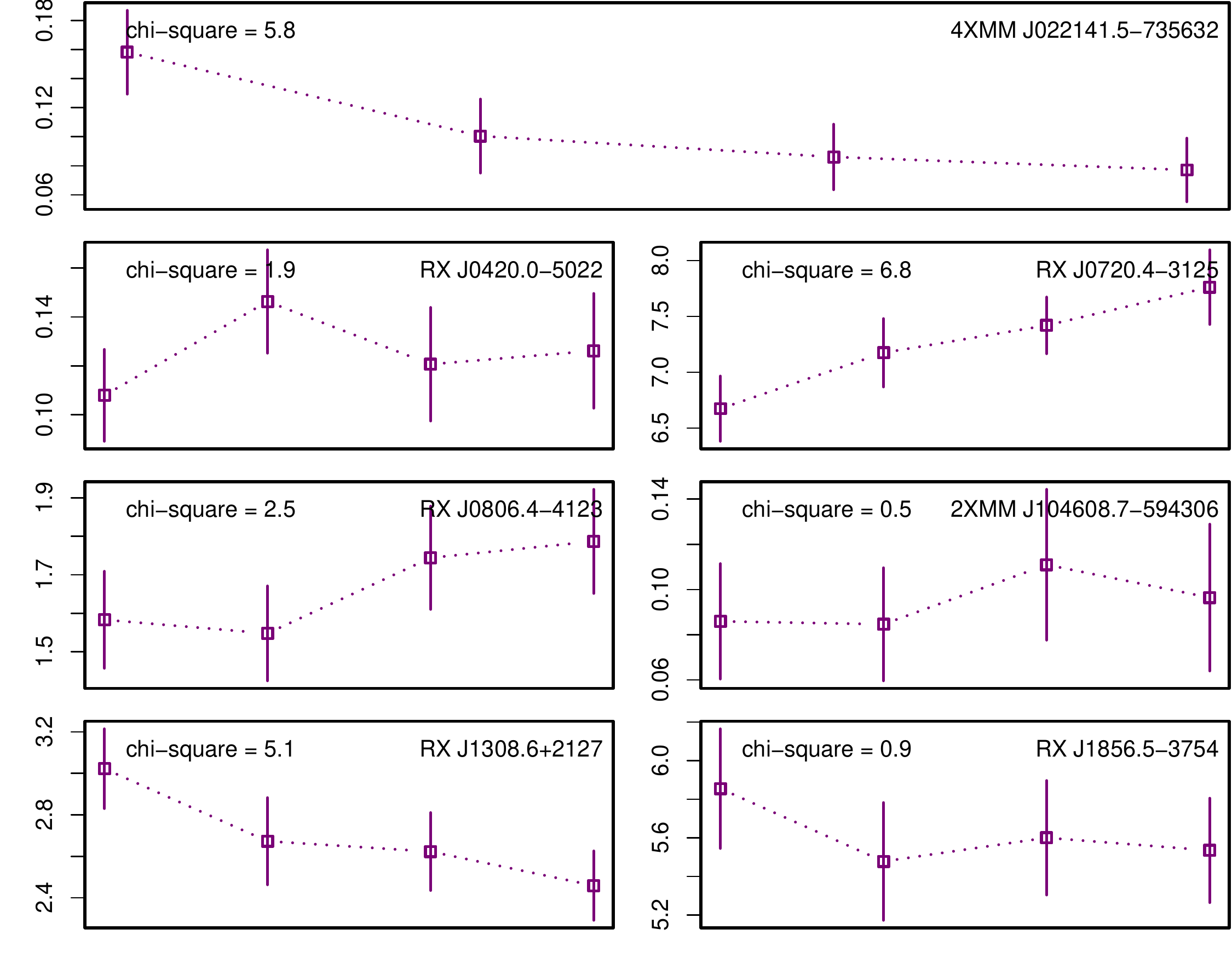}
\caption{Count rates of thermally emitting INSs in eRASS1-4 (TM8, $0.2-1$\,keV). A chi-square test for constancy is formally rejected for \jztto, \magzf, \magzs, \magze, and \magot. All rates are consistent within $2\sigma$.
\label{fig_ratesm7}}
\end{center}
\end{figure*}

We investigated whether similar variations are observed in the survey data of other thermally emitting INSs. We retrieved the c020 pipeline processed sky tiles centred on the positions of \magzf, \magzs, \magze, \magot, \magoe, and \carINS\ (Table~\ref{tab_erassm7}), and analysed them consistently with the procedure described here for \insc. The $0.2-1$\,keV count rates of the sources from PSF fitting as a function of eRASS epoch can be seen in the diagram of Fig.~\ref{fig_ratesm7}, with $1\sigma$ errors. A chi-square test for constancy at p-value 0.1 is formally rejected for all sources except \magoe\ and \carINS\ ($\chi^2$ values within 1.9 and 6.8 above the critical value of $\chi^2_{\rm cr}\sim0.58$ for 3 d.o.f.). All rates are consistent within $2\sigma$.

To fit the individual eRASS1--4 spectra of the six INSs we assumed simple absorbed blackbody models. The fit results are in general agreement with those previously reported in the literature. We found up to $2\sigma$ variations in flux and deviations of up to $23\%$ in $kT$, suggesting that the level of intra-eRASS variability observed for the INS candidate \insc\ is typical and probably a result of low-count statistics. 
\section{Variable elemental abundances\label{sec_additspec}}
In Table~\ref{tab_varabs} we show the results of two additional spectral fits with the variable absorption model \texttt{tbvarabs} in XSPEC, which were not included in Table~\ref{tab_spec} for conciseness.
Model (A) is a single-temperature constant blackbody with enhanced oxygen abundance with respect to solar, $Z_{\rm O}=(5.4_{-1.9}^{+2.4})\times Z_{\odot}$; it consists of a significant improvement with respect to model (1) in Table~\ref{tab_spec} (F-test probability 0.003). Elemental abundances other than oxygen were either insensitive or unconstrained by the fit. Similarly to the improvement found for the multi-component models (2) and (3), the best-fit parameters of the blackbody component do not significantly ($>2\sigma$) differ from those of model (1).

In model (B) we tested the possibility that the source is extragalactic, motivated by its projected location in the Magellanic Bridge and $\nh$ in excess of the Galactic value from \citet{1990ARA&A..28..215D}, $N_{\rm H}^{\rm DL}=3.8\times10^{20}$\,cm$^{-2}$. As discussed in Section~\ref{sec_disc}, the hydrogen column density integrated in the line-of-sight includes the contribution of both the Milky Way and the Magellanic Clouds; ideally, these should be uncoupled to get a reliable estimate of the foreground $\nh$ in our Galaxy \citep[e.g.][]{2022A&A...661A..37S}. 
Assuming that the source is located at the distance of the Small Magellanic Cloud (SMC), $d_{\rm SMC}=60$\,kpc, we thus tested a model with two absorption components. We set $N_{\rm H}^{\rm Gal}\equiv N_{\rm H}^{\rm DL}$ whereas $N_{\rm H}^{\rm SMC}$ was free to vary. For the latter, the elemental abundances of \citet{1989GeCoA..53..197A} were initially fixed to a sub-solar value of 0.2, more typical of that of the SMC \citep{1992ApJ...384..508R}, then allowed to vary to search for an improvement in the fit statistic. The best solution, with $N_{\rm H}^{\rm SMC}=3.4_{-2.0}^{+2.2}\times10^{20}$\,cm$^{-2}$ and $kT=67.2\pm2.9$\,eV, is obtained when oxygen is overabundant with respect to solar, $Z_{\rm O}=(4_{-2}^{+8})\times Z_\odot$; we note that this is essentially the same fit result as that of model (A). In Table~\ref{tab_varabs} we list the corresponding emission radius of the model normalised to $d_{\rm SMC}$, $R_{\rm em}^{\rm SMC}=340_{-50}^{+70}$\,km.
\begin{table}[t]
\small
\caption{Results of spectral fitting with variable absorption.
\label{tab_varabs}}
\centering
\begin{tabular}{@{}lcrrrrr@{}}
\hline\hline
 & NHP\tablefootmark{a} & $C$\,(d.o.f.) & $\nh$\,\tablefootmark{b} & \multicolumn{2}{c}{Model parameters} & $F_X$\,\tablefootmark{c}\\
\hline
\multicolumn{4}{l}{\texttt{tbvarabs*bbodyrad}} & $kT$ & $R_{\rm em}$\,\tablefootmark{d} & \\
 & & & & (eV) & (km) & \\
\hline
A\tablefootmark{$\star$} & 70 & $153\,(158)$ & $5.3(9)$ & $67(3)$ & $5.6_{-0.9}^{+1.2}$ & $4.3_{-0.6}^{+0.7}$\\
B\tablefootmark{$\dagger$} & 70 & $154\,(158)$ & $3.4_{-2.0}^{+2.2}$ & $67.2(2.9)$ & $340_{-50}^{+70}$ & $4.4_{-0.5}^{+0.7}$\\
\hline
\end{tabular}
\tablefoot{Errors are $1\sigma$ confidence levels. Models fitted to the data: (A) single-temperature blackbody and (B) ``extragalatic'' model with \citet{1989GeCoA..53..197A} abundances, fixed $N_{\rm H}^{\rm Gal}\equiv3.8\times10^{20}$\,cm$^{-2}$ \citep{1990ARA&A..28..215D}, and best-fit $N_{\rm H}^{\rm SMC}$ listed above.
\tablefoottext{a}{Null-hypothesis probability in percentage that the observed data are drawn from the model.}
\tablefoottext{b}{The column density is in units of $10^{20}$\,cm$^{-2}$.}
\tablefoottext{c}{The unabsorbed flux is in units of $10^{-13}$\,erg\,s$^{-1}$\,cm$^{-2}$ ($0.2-12$\,keV).}
\tablefoottext{d}{The blackbody emission radius at infinity is computed assuming a distance of $1$\,kpc (A) and 60\,kpc (B).}
\tablefoottext{$\star$}{The best-fit abundance of oxygen is $Z_{\rm O}=(5.4_{-1.9}^{+2.4})\times Z_{\odot}$.}
\tablefoottext{$\dagger$}{The best-fit abundance of oxygen is $Z_{\rm O}=(4_{-2}^{+8})\times Z_{\odot}$.}
}
\end{table}
\end{appendix}
\end{document}